\documentclass{aa}
\usepackage{natbib}
\bibpunct{(}{)}{;}{a}{}{,} 
\input{bib.def}

\usepackage{graphicx}
\usepackage{txfonts}

\usepackage{ulem}
\begin{document} 

   \title{The Sun at millimeter wavelengths}
   \subtitle{I. Introduction to ALMA Band~3 observations}

   \author{Sven Wedemeyer$^*$ \inst{1,2}
          \and
  Mikolaj Szydlarski  \inst{1,2}
          \and
  Shahin Jafarzadeh  \inst{1,2}
          \and 
  Henrik Eklund  \inst{1,2}
          \and 
  Juan Camilo Guevara Gomez \inst{1,2}
          \and
  Tim Bastian \inst{3}
          \and
  Bernhard Fleck \inst{4}
          \and
  Jaime de la Cruz Rodriguez \inst{5}
          \and
  Andrew Rodger \inst{6}
          \and
  Mats Carlsson \inst{1,2}
  }
   
  \authorrunning{Wedemeyer {et~al.}}
   \institute{Rosseland Centre for Solar  Physics, University of Oslo, Postboks 1029 Blindern, N-0315 Oslo, Norway \\ \email{sven.wedemeyer@astro.uio.no}
            \and
            Institute of  Theoretical Astrophysics, University of Oslo, Postboks 1029 Blindern, N-0315 Oslo, Norway 
            \and
            National Radio Astronomy Observatory (NRAO), 520 Edgemont Road, Charlottesville, VA 22903, USA
            \and 
            ESA Science Operations Department, c/o NASA Goddard Space Flight Center, Greenbelt, MD 20771, USA
            \and
            Institute for Solar Physics, Dept. of Astronomy, Stockholm University, Albanova University Center, SE-10691 Stockholm, Sweden         
            \and
            SUPA, School of Physics and Astronomy, University of Glasgow, Glasgow G12 8QQ, United Kingdom} 
 

   \date{Received --- ; accepted --- }

\abstract{The Atacama Large Millimeter/sub-millimeter Array (ALMA) started regular observations of the Sun in 2016, first offering receiver Band 3 at wavelengths near 3\,mm (100 GHz) and Band 6 at wavelengths around 1.25\,mm (239 GHz). 
}
{Here, we present an initial study of one of the first ALMA Band~3 observations of the Sun with the aim to characterise the diagnostic potential of brightness temperatures measured with ALMA on the Sun. 
}
{The observation covers a duration of 48\,min at a cadence of 2\,s targeting a Quiet Sun region at disk-centre. 
Corresponding time series of brightness temperature maps are constructed with the first version of the Solar ALMA Pipeline (SoAP) and compared to simultaneous observations with the Solar Dynamics Observatory (SDO). 
%
%
}
{The angular resolution of the observations is set by the synthesized beam, an elliptical Gaussian that is approximately $1.4\arcsec\times 2.1\arcsec$ in size.
The ALMA maps exhibit network patches, internetwork regions and also elongated thin features that are connected to large-scale magnetic loops as confirmed by a comparison with SDO maps. 
The ALMA Band 3 maps correlate best with the SDO/AIA 171\,\AA, 131\,\AA\ and 304\,\AA\ channels in that they exhibit network features and, although very weak in the ALMA maps, imprints of large-scale loops. 
A group of compact magnetic loops is very clearly visible in ALMA Band 3. The brightness temperatures in the loop tops reach values of about  8000-9000\,K and in extreme moments up to 10\,000\,K.
}
{ALMA Band 3 interferometric observations from early observing cycles already reveal temperature differences in the solar chromosphere. 
The weak imprint of magnetic loops and the correlation with the 171, 131, and 304 SDO channels suggests though that the radiation mapped in ALMA Band~3 might have contributions from a larger range of atmospheric heights than previously assumed but the exact formation height of Band 3 needs to be investigated in more detail.
The absolute brightness temperature scale as set by Total Power measurements remains less certain and must be improved in the future. 
Despite these complications and the limited angular resolution, ALMA Band 3 observations have large potential for quantitative studies of the small-scale structure and dynamics of the solar chromosphere. 
}


   \maketitle
%
\begin{figure*}[tp!]
\sidecaption
\includegraphics[width=12cm]{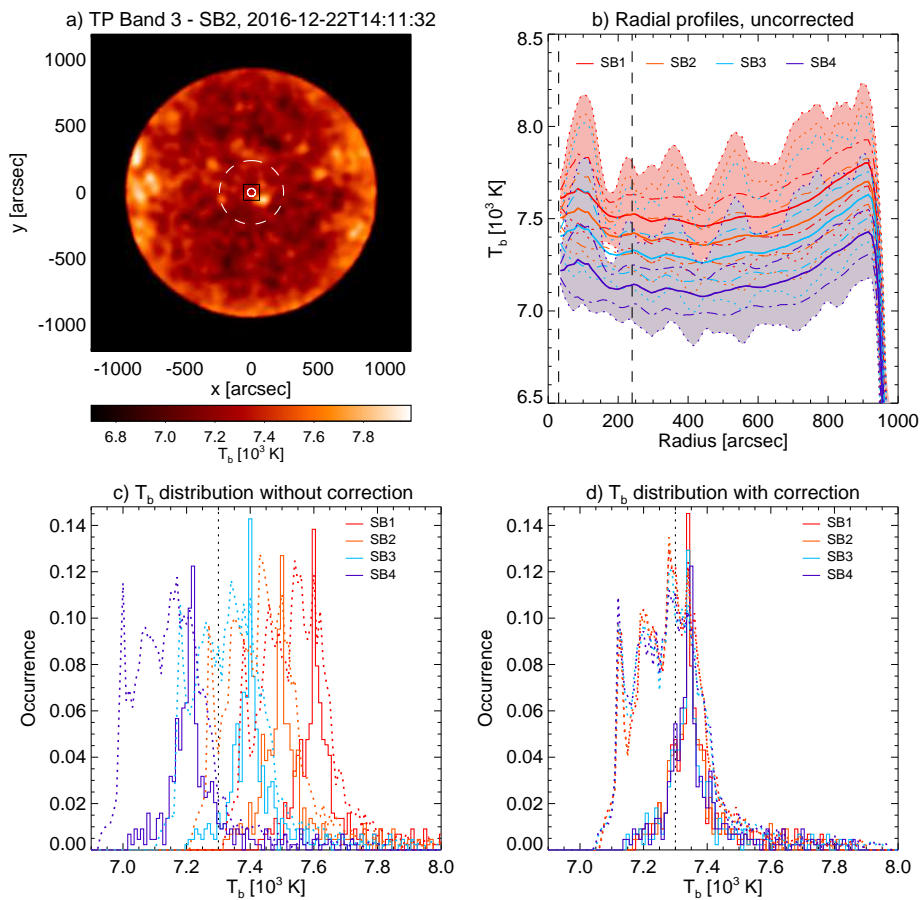}
\caption{Brightness temperature distribution in the Total Power Band~3 maps completed at \mbox{2016-12-22T14:11:32.}  
\textbf{a)}~Total Power map for SB2 (correction following \citet{2017SoPh..292...88W}  applied, i.e., mean value in central region rescaled).  
The white solid circle in the middle indicates the primary beam of a TP antenna at 3.0\,mm and with it the field-of-view (FOV) of the interferometric observation. The dashed circle has a radius of 240\,\arcsec{}  and marks the region that is considered for the correction of the absolute brightness temperature scale. The black square marks the central region with 120\,\arcsec\,$\times$\,120\,\arcsec.  
\textbf{b)}~Uncorrected radial profiles for all sub-bands in rings of 10\,\arcsec{}  width. The lines for each sub-band (see color legend at the top) represent the radial average (solid), average plus/minus the standard deviation (dot-dashed), and the 1$^\mathrm{st}$ and  99$^\mathrm{th}$ percentile (dotted). The shaded areas cover the value ranges between the percentiles for SB1 (red) and SB4 (blue).  
The bottom row shows histograms for the brightness temperature distribution within the black square (solid histograms)  and within the 
white dashed circle (dotted histograms, scaled by a factor 3) for all sub-bands \textbf{c)}~without  and \textbf{d)}~with correction to the reference value suggested by \citet{2017SoPh..292...88W} (vertical black dotted line).
}\label{fig:almatp}
\end{figure*}
\section{Introduction}

The Atacama Large Millimeter/submillimeter Array (ALMA) provides new diagnostic possibilities to probe the chromosphere of the Sun  
 at high spatial, temporal, and spectral resolution
\citep[see][and references therein]{2002AN....323..271B,2016Msngr.163...15W, 2016SSRv..200....1W,2017A&A...601A..43L,2018Msngr.171...25B}.
In principle, observing at millimeter wavelengths has the advantage that the radiation is formed under conditions of local thermodynamic equilibrium (LTE) and therefore provides a more direct measure of local gas temperatures in the chromosphere than 
other commonly used diagnostics at shorter wavelengths, such as optical and UV wavelengths, that are not in LTE. 
The comparatively long millimeter and submillimeter wavelengths have historically had the disadvantage of a correspondingly lower angular resolution, relying largely on single dish observations (e.g., Bastian et al. 1993, Lindsey et al. 1995, and references therein). Interferometric techniques, using an array of antennas, offer the means of observing the Sun with high angular resolution. These were explored in the 1990s (e.g., Kundu et al. 1993) and 2000s (White et al. 2006) using small arrays. ALMA is the largest and most ambitious array ever built to observe celestial phenomena at millimeter and submillimeter wavelengths, including the Sun. ALMA offers the potential of unlocking this new diagnostic tool for high-resolution studies of the solar chromosphere. 
An overview of potential science cases with ALMA is given by \citet{2016SSRv..200....1W}, whereas Cycle 4 capabilities are described by \citet{2017SoPh..292...88W} and \citet{2017SoPh..292...87S}.
While first regular ALMA observations of the Sun were only offered in Cycle~4 with a first solar campaign in December 2016, 
earlier observations from Commissioning and Science Verification (CSV) campaigns have been made publicly available. 
Both regular and CSV data are already used in publications: \citet[][]{2017A&A...605A..78A,2017ApJ...845L..19B,2017ApJ...841L...5S,2018A&A...613A..17B,2018A&A...619L...6N,2018ApJ...863...96Y,2019A&A...622A.150J,2019ApJ...877L..26L,2019ApJ...881...99M,2019ApJ...875..163R,2019ApJ...871...45S,2019arXiv191203480P,2019arXiv191209886S}.

Interferometric observations of a dynamic source like the Sun and the reliable reconstruction of corresponding image series are  challenging tasks. 
As a next step, in order to further develop and characterise ALMA's diagnostic capabilities, the available observations have to be thoroughly studied and compared to other diagnostics. 
Here, we present and analyse observations with ALMA~Band~3 at wavelengths around 3\,mm from December~2016 (Cycle~4), which were among the first regular observations of the Sun with ALMA. 
The aim of the results presented here is to illustrate the potential, limitations, and challenges of studying the small-scale structure and dynamics of the solar atmosphere with ALMA~Band~3. 
The technical details of the observations are described in Sect.~\ref{sec:material} and the results of the data analysis in Sect.~\ref{sec:results}. 
Discussion and conclusions are provided in Sects.~\ref{sec:disc} and \ref{sec:conc}, respectively.

\section{Observations} 
\label{sec:material}
\label{sec:observ}


\subsection{Solar observation in Band 3}

The Band~3 observations discussed in this article were carried out on December 22, 2016 
from 14:22UT - 15:07UT. 
ALMA observations of the Sun currently comprise both interferometric observations of a specific target and full disk maps made with  ALMA total power (TP) antennas. 
For the interferometric observations, an array in configuration C40-3 was used, which included a total of 52~antennas from the
12-m Array as well as the 10 fixed 7m antennas of the Atacama Compact Array. 
The resulting array has baselines ranging from 9.1\,m to 492.0\,m resulting in a nominal angular resolution of 1.56\,\arcsec{} and a Maximum Recoverable Scale (MRS) of 68\,\arcsec. 
%
In addition to the interferometric observations, ALMA has up to 4 specially designed Total Power (TP) antennas that can perform rapid scans of the whole  disk of the Sun \citep{2017SoPh..292...88W}. 
For the observations analysed here, three TP antennas were available for fast-scan mapping.  
The column of precipitable water vapour (PWV) in Earth's atmosphere during the observation was  1.60\,mm.

Because of an operational glitch, the interferometric array did not point at and track the intended target region but instead 
re-centred on \mbox{[x,y]\,=\,[0”,0”]} in helioprojective coordinates repeatedly during the observation.
As a result, the observed disk-centre Quiet Sun region is slowly drifting through the ALMA field-of-view (FOV) because the telescope pointing did not track solar rotation (see Sect.~\ref{sec:intprocess}). 
The Band~3 observing sequence consists of 4~scans with a duration of $\sim$10\,min each. These scans are separated by calibration breaks of $\sim$2.4\,min. 
The observations were carried out with a cadence of 2\,s - the highest possible in Cycle~4. 

In Cycle 4, ALMA Band~3 was set up for solar observations in 4 spectral windows (hereafter referred to as sub-bands) around a central frequency of 100\,GHz. These sub-bands, which we refer to as sub-bands 1 -- 4 (abbreviated SB1 -- SB4) with increasing frequency, are centered on 93\,GHz (SB1), 95\,GHz (SB2), 105\,GHz (SB3), and 107\,GHz (SB4), corresponding to wavelengths of 3.224\,mm, 3.156\,mm, 2.855\,mm, and 2.802\,mm. 
Each sub-band has a total bandwidth of 2\,GHz (with the central 1.875\,GHz being retained), which results in two pairs of neighbouring sub-bands (SB1-SB2 and SB3-SB4) with a central gap. 

The three available TP maps  were completed at 14:23UT, 14:36UT, and 14:49UT. 
A complete scan in Band~3 took between 12.6\,min and 12.9\,min, which includes calibration. 
The net time for scanning the solar disk in a double-circle pattern is 5\,min. 
%
The TP maps thus cover most of the interferometric observation and can be used for combining the interferometric and TP data, which results in absolute brightness temperatures. 
The TP map for SB4 for the first scan (from 14:11UT to 14:23UT) is shown in Fig.~\ref{fig:almatp}a. 
Please refer to Sect.~\ref{sec:app_tp} for background information regarding the TP observations.

\subsection{Interferometric data processing} 
\label{sec:intprocess}

\paragraph{Approach for the Band 3 data set.} 
The calibrated ALMA data   were downloaded from the ALMA Archive and  further processed with the Solar ALMA Pipeline 
(SoAP, \citeauthor{soap_inprep} in prep.\footnote{SoAP is an initiative of the SolarALMA project in Oslo in collaboration with the international solar ALMA development team.}) based on the Common Astronomy Software Applications (CASA\footnote{CASA:  \texttt{http://casa.nrao.edu}}) package. 
Please note that solar observing is currently still a non-standard mode. Solar data are therefore not processed with the official ALMA pipeline. 
Instead, SoAP is used for this publication. 
Please refer to Sect.~\ref{sec:app_intimg} in the appendix for more information on interferometric image reconstruction.

Unique to data from December 2016 are the complications arising from the erroneous pointing and tracking, which resulted in the instrument phase tracking center repeatedly being re-pointed to the apparent center of the solar disk, resulting in a slow drift of the FOV with intermediate jumps. 
It was therefore necessary to correct for the effect of the Sun's rotation during the course of the observation. To do so, a time series of Band 3 images at 2s cadence was constructed. The image processing required for each snapshot image includes several important steps: first, the ALMA PSF (the ``dirty beam'') is deconvolved from the image data (the ``dirty map'') through application of the multi-scale (multi-frequency) CLEAN algorithm \citep{2011A&A...532A..71R} as implemented  in CASA. 
Here, all interferometric information from the four sub-bands is used to produce one continuum  image (referred to as ``full-band map'') for each time step (see also Sect.~\ref{sec:app_intimg}). 
Second, the image data are corrected for the effect of the primary beam (see Sect.~\ref{sec:fov}). Third, the interferometric data are combined (also called ``feathered'') with the TP map in order to add an (DC brightness) offset and thus the absolute brightness temperature scale corresponding to zero-spacing information to the reconstruction (Sect~\ref{sec:tp}). 
Finally, the ALMA Band 3 maps are co-aligned with observational data from other observatories (Sect.~\ref{sec:coobserv}). The apparent drift of snapshot images in time due to solar rotation was then corrected by cross-correlating consecutive 2\,s images with a reference image, where the first frame in the times series was taken as the reference. 
The resulting time sequence represents true snapshot imaging at 2\,s cadence with no temporal averaging. 
We would like to emphasise that self-calibration for a short time window is the recommended approach  but that self-calibration  resulted in too aggressive corrections and loss of information on small spatial scales for December 2016 data suffering from pointing errors. 
A detailed description of the data processing with SoAP will be provided in a forthcoming publication \citep{soap_inprep}.

\subsection{Interferometric field-of-view} 
\label{sec:fov}

For the Band~3 data discussed here, the FWHM beam width varies from 67.5\,\arcsec{} for SB1 to 60.0\,\arcsec{}  for SB4 with 63.8\,\arcsec{} for the band centre frequency. 
Please refer to Sect.~\ref{sec:app_fov} in the appendix for general background information. 
Since, as a result of the primary beam taper the source brightness decreases with distance from the beam axis while the noise stays constant, the signal-to-noise ratio declines as function of distance from the beam axis. To correct for the primary beam taper, the image data are divided by the relevant Gaussian (unit maximum) out to some user-specified threshold level where the SNR remains significant. 
The resulting FOV for interferometric ALMA images is therefore set by the wavelength (or frequency) and the chosen threshold for the primary beam. 
A threshold of 0.3 is a reasonable but generous choice and results in diameters of the FOV from 89\,\arcsec{}  (SB1) to 79\,\arcsec{}  (SB4).
In the particular case of the data from December 2016, problems with the pointing and resulting coordinate jumps led to a reduction of the final FOV once the data had been corrected for solar rotation.
The resulting FOV of these maps was set to a diameter of 65.6\,\arcsec{}, which corresponds to effective Gaussian thresholds of 0.52 for SB1 and 0.44 for SB4, respectively. 

\subsection{Synthesized beam} 
\label{sec:observbeam}

We define the beam representative  for the observations considered here (which determines the angular resolution) as 
the time-averages of the major axis, minor axis, and position angle. 
The resulting representative beam corresponds to the band-average frequency of 100\,GHz and has a major axis of 2.10\,\arcsec{}  (full-width-half-maximum, FWHM ), a minor axis  of 1.37\,\arcsec{}  (FWHM ) and a position angle of 68.0\,deg (see Fig.~\ref{fig:almapsf}). 
The beam for the time step at 2016-12-22 14:42:04UT comes closest to the representative beam in terms of size. 
During the 48\,min covered during the observation with the Sun moving on the sky, the major axis shrank by $\sim$7\,\% (see Fig.~\ref{fig:almapsf}c), whereas the minor axis stayed almost constant and the position angle increased by less than 2~degrees  (see Fig.~\ref{fig:almapsf}d). 
The changes of the beam must be taken into account for a meaningful interpretation of the resulting data. 
Please refer to Sect.~\ref{sec:app_beam} for more details.

\begin{figure*}[ht!]
\begin{center}
\includegraphics[width=\textwidth]{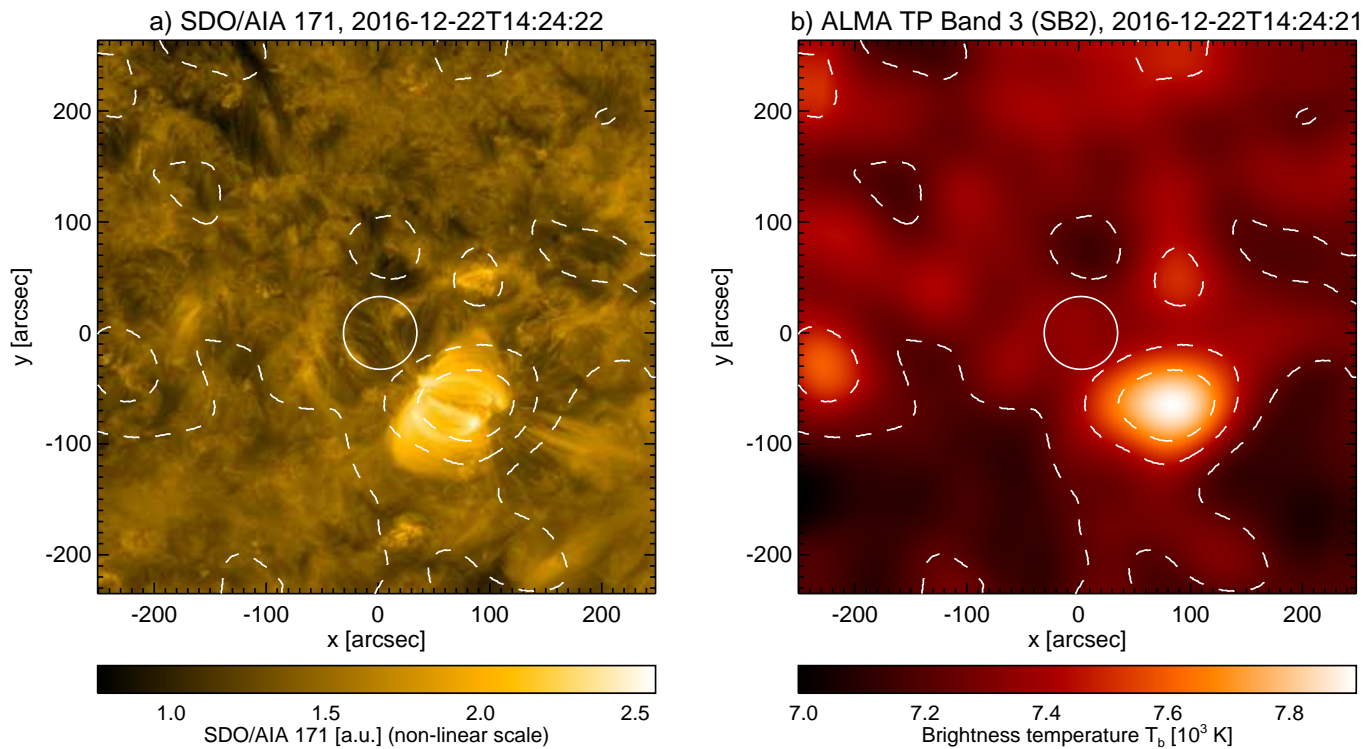}
\end{center}
\caption{Context maps for the surroundings of the interferometric field-of-view (FOV, white solid circle). 
\textbf{a)}~SDO/AIA 17.1\,nm. (The SDO data are post-processed (level~2) in order to increase the visibility of the atmospheric structure.)
\textbf{b)}~ALMA TP Band~3 (sub-band~2). The white dashed lines are contours for the ALMA data. The displayed data range for the SDO map is limited to 20\,\% of the maximum in order to make dimmer structures more visible. The beam size of the ALMA map is comparable to the shown interferometric FOV.} 
\label{fig:almaoverview}
\end{figure*}

\subsection{Absolute temperatures based on Total Power maps}
\label{sec:tp}

\citet{2017SoPh..292...88W} suggest that, until systematic errors in the dual-load calibration scheme are fully understood and resolved, ALMA Band 3 TP maps should be scaled to a prescribed value of 7300\,K. 
The TP maps for the data presented here were produced for each sub-band and calibrated using the dual-load approach described by \citet{2017SoPh..292...88W}  as implemented in CASA\footnote{The CASA versions of the dual-load calibration scripts were produced in connection with a meeting of the International Team 387 funded by the International Space Science Institute (ISSI, Bern, Switzerland).}.   
Please note that, for Band~3, \citet{2017SoPh..292...88W} recommend to use the average over the inner square region  with a size 120\,\arcsec\,$\times$\,120\,\arcsec{} (black square in Fig.~\ref{fig:almatp}a), whereas the CASA script provided with the TP data uses the average over the central region of the solar disk with a radius of 240\,\arcsec{} (40~pixels, see dashed white circle in Fig.~\ref{fig:almatp}a). 
The  histograms in Fig.~\ref{fig:almatp}c show the absolute brightness temperatures for the two different regions for the different sub-bands. 
The average brightness temperatures for the inner 120\,\arcsec\,$\times$\,120\,\arcsec{} in the 
TP map used here (\mbox{2016-12-22T14:11:32}) are   
7635\,K (SB1), 7529\,K (SB2), 7434\,K (SB3), and 7247\,K (SB4) 
as compared to the corresponding average values for the circular region with radius of 240\,\arcsec:
7559\,K (SB1), 7454\,K (SB2), 7357\,K (SB3), and 7171\,K (SB4). The values for the inner 120\,\arcsec\,$\times$\,120\,\arcsec{} region are thus 75-77\,K higher than the larger circular region. 
It should be noted that the 120\,\arcsec\,$\times$\,120\,\arcsec{} region is relatively small considering the width of the primary beam ($\sim60$\arcsec), resulting in poor statistics and susceptibility to untypical brightness temperatures. 
As a consequence, the histograms for the central square 120\,\arcsec\,$\times$\,120\,\arcsec{}  in Fig.~\ref{fig:almatp}c-d are much narrower as compared to the histograms of the inner region with a radius of 240\,\arcsec. 
It should also be noted that the common procedure is to use only one TP sub-band (typically SB2) for determining the offset and combination with the interferometric data, which thus ignores data from the other three sub-bands. 
%

\begin{figure*}[hp!]
\begin{center}
\includegraphics[width=\textwidth]{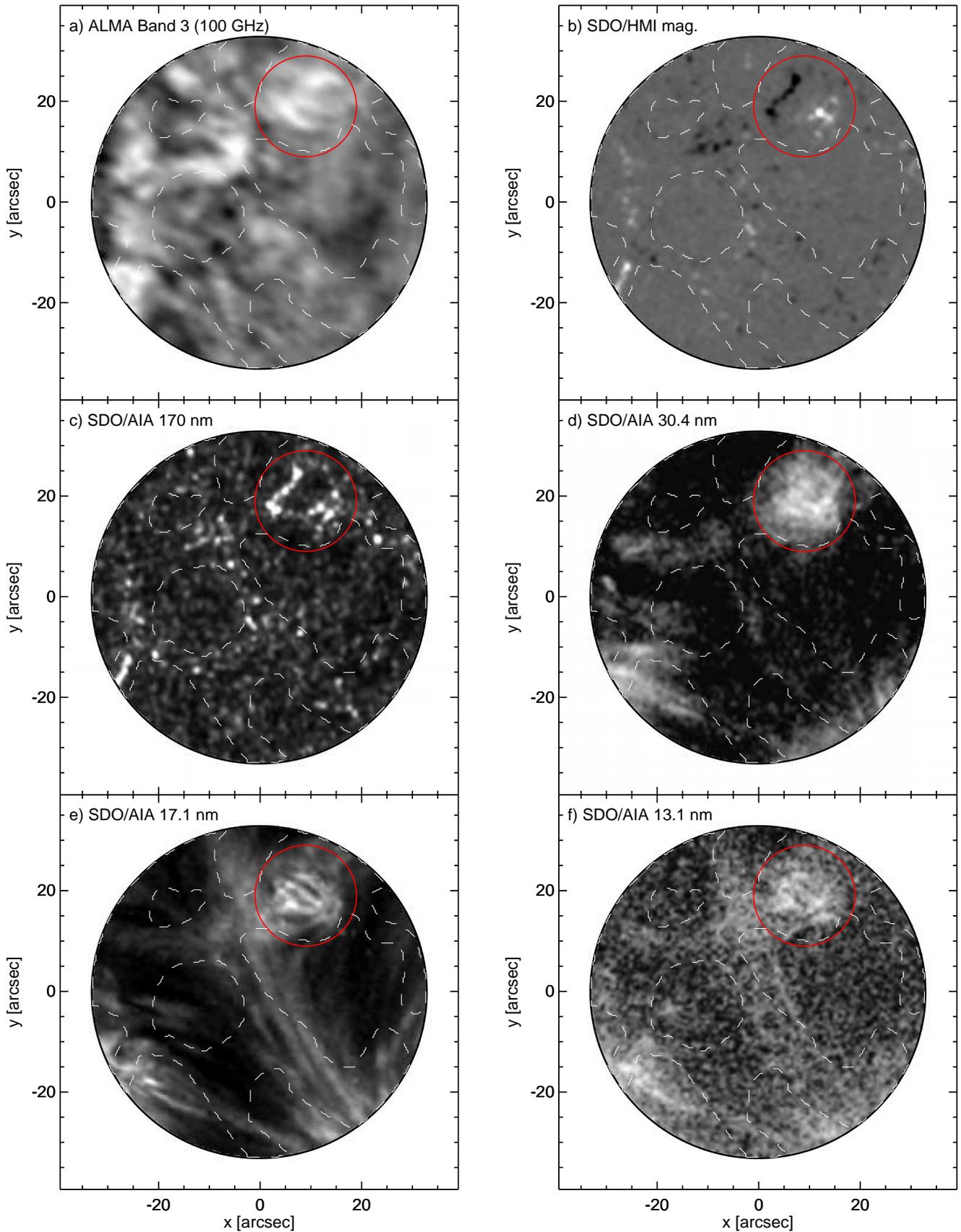}
\end{center}
\vspace*{-8mm}
\caption{
\textbf{a)}~Representative ALMA Band 3 image at 2016-12-22 14:22:52UT in comparison to co-aligned SDO maps:  
\textbf{b)}~SDO/HMI magnetogram, 
\textbf{c)}~SDO/AIA 170\,nm, 
\textbf{d)}~SDO/AIA 30.4\,nm, 
\textbf{e)}~SDO/AIA 17.1\,nm,  
\textbf{f)}~SDO/AIA 13.1\,nm.  
The SDO data are post-processed (level~2) and re-scaled  in order to increase the visibility of the atmospheric structure. 
The network/internetwork mask is overplotted with dashed contours. 
The red circle marks a region with a compact group of magnetic loops.
}\label{fig:almasdo}
\end{figure*}

The aforementioned mean values, and also the distribution peak temperatures, are highest for SB1 and lowest for SB4, consistent with the expectation that SB1 is formed higher in the solar atmosphere and that the average gas temperature in the mapped layers is monotonically increasing, as, e.g., reflected by the classic semi-empirical models of \citet{Vernazza_1981ApJS...45..635V}. 
On the other hand, the differences between the peak temperatures do not scale according to the sub-band frequencies and are not grouped accordingly into two pairs, suggesting  offsets in the brightness temperatures of possibly on the order of 100\,K.  
The radial brightness temperature averages in  Fig.~\ref{fig:almatp}b  show the same differences between the sub-bands and thus the same order. 
The standard deviation is for all sub-bands between 100\,K and 150\,K for radii between 150\,\arcsec{} and 900\,\arcsec, which is in line with the statistical uncertainty found by \citet{2017SoPh..292...88W}. 
Following the re-scaling procedure recommended by \citet{2017SoPh..292...88W} for all interferometric sub-bands separately would then shift the distributions of all sub-bands to roughly the same peak value (see Fig.~\ref{fig:almatp}d). While correcting offsets between the sub-bands, this procedure would also remove brightness temperature differences between the sub-bands that are connected to slightly different formations heights and the average temperature increase in the chromosphere. The resulting corrected sub-band differences are misleading in the sense that they do not reflect the true temperature gradients in the solar atmosphere.

We note that, for the observation presented here, there is a bright feature  in the inner region (dashed circle in Fig.~\ref{fig:almatp}a, see also Fig.~\ref{fig:almaoverview}) that becomes a strong plage or enhanced network region of opposite polarity in the days following the observation. 
Excluding the bright feature would change the average value for the inner region with radius 240\arcsec{} for SB2 from originally 7454\,K to 7438\,K. 
This feature alone thus produces a 16\,K shift in the absolute brightness temperature scale.
Further improvements to the calibration removing such effects are desirable.

For the data presented here, we strictly follow the  procedure implemented in the officially provided CASA script and rescale the average over the (unaltered) central region with a radius of 40~pixels (corresponding to 240\arcsec) in the TP~SB2 map to the recommended reference value of 7300\,K.
The average value in the original SB2 map is 7454\,K and thus only 154\,K higher than the recommended value, resulting in an applied scaling factor of 0.979 for the whole map.
For the moment, significant uncertainties of the TP maps and thus the absolute brightness temperatures remain but will be reduced by future improvements of the calibration procedure. 

\subsection{Final data product and co-observations with SDO} 
\label{sec:coobserv}

Post-processing with SoAP produced one time series of 1200 full-band (continuum) ALMA maps  with a cadence of 2\,s divided into 4~scans of \mbox{$\sim10$\,min} duration (300~maps each) and intermediate \mbox{$\sim2.4$\,min} breaks. 
As mentioned in Sect.~\ref{sec:fov}, the FOV of these maps was limited to a diameter of 65.6\,\arcsec{} to ensure that each pixel in the FOV has data for all time steps. 
The full-disk TP maps for each of the three TP scans are also available for the analysis. 
Cotemporaneous observations with the Solar Dynamics Observatory  \citep[SDO;][]{2012SoPh..275....3P} 
recorded with the Atmospheric Imaging Assembly 
\citep[AIA][]{2012SoPh..275...17L} 
and the Helioseismic and Magnetic Imager \citep[HMI;][]{2012SoPh..275..207S} instrument are used here. 
The ALMA and SDO images have been co-aligned for the whole duration of the observation covering $\sim$47.6\,min including ALMA intermediate calibration breaks. 
A representative timestep is presented in Fig.~\ref{fig:almasdo}a. 
Movies for the whole time series are provided as online material: (i)~The ALMA FOV alone and (ii) in comparison to SDO channels. For the movies, additional boxcar averaging with a window of 20\,sec is applied \citep[cf.][]{1994ApJ...430..413S}.

\subsection{Data mask} 

\begin{figure}[tp!]
\begin{center}
\includegraphics[width=\columnwidth]{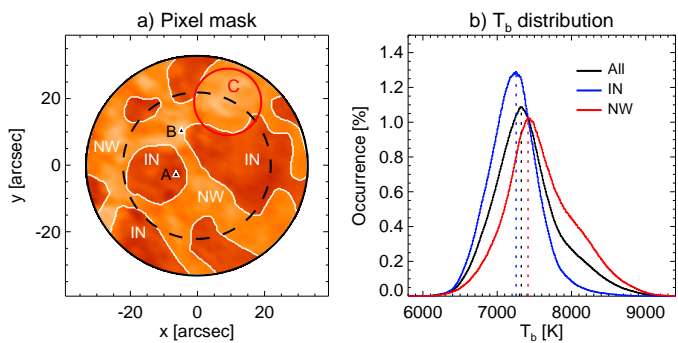}
\end{center}
\vspace*{-5mm}
\caption{\textbf{a)} 
The pixel mask distinguishing between network (NW) and internetwork (IN) pixels (white lines separate NW and IN). The black dashed circle marks the inner region with a radius of 22\,\arcsec{}. The red circle shows the location of a group of compact loops (labeled C). 
The triangles mark two locations (A and B) for which profiles are shown in Fig.~\ref{fig:almaprof}.
\textbf{b)}~Brightness temperature distributions in the inner regions of the FOV (radius $r \leq 22$\,\arcsec{})   over the whole observing time period. 
All pixels (black) compared to internetwork (blue) and network pixels (red) for the full band maps.}\label{fig:almahist}
\end{figure}

The FOV (see Fig.~\ref{fig:almasdo}a, see also Fig.~\ref{fig:almaoverview}) contains a Quiet Sun region with a mixture of magnetic network and internetwork patches. 
In order to distinguish between Quiet Sun internetwork and network pixels, a data mask is constructed based on a combination of time-averaged  maps in SDO/AIA\,1700 and SDO/AIA\,1600 and saturated SDO/HMI magnetograms and the band-averaged ALMA maps. 
The time-averages include the whole observing period. 
The final mask is shown in Fig.~\ref{fig:almahist}a.

\begin{figure}[t!]
\begin{center}
\includegraphics[width=\columnwidth]{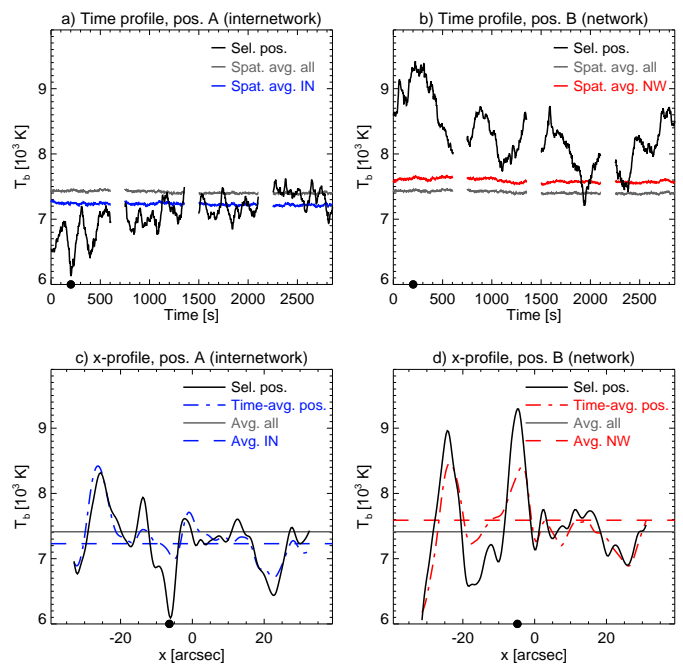}
\end{center}
\vspace*{-4mm}
\caption{ALMA brightness temperatures for the two selected positions  A~(internetwork) and   B~(network), which are marked in Fig.~\ref{fig:almahist}. 
The upper row shows the temporal evolution for  
\textbf{a)}~the internetwork position~A and \textbf{b)}~the network position~B. 
The dot on the abscissa marks the time step shown in  Fig.~\ref{fig:almasdo}. 
The temporal evolution of the average over all pixels in the inner 
region  (radius $r \leq 22$\,\arcsec, grey line) and over the contained internetwork (IN, blue line) and network (NW, red line) pixels is plotted for comparison.
Profiles along the x-axis are shown in the lower row for the same time step for \textbf{c)}~position~A (internetwork) and \textbf{d)}~position~B (network).  
The selected spatial position in Fig.~\ref{fig:almahist}a is marked with a dot on the abscissa. 
For comparison, the time-averages at the selected positions (blue/red dot-dashed lines), the averages over all time steps for all pixels in the inner region (grey solid line) and for the internetwork and network pixels (blue/red dashed lines), respectively, are shown.
}\label{fig:almaprof}
\end{figure}

\section{Results} 
\label{sec:results}

\subsection{Atmospheric structure observed with ALMA and SDO} 
\label{sec:observstruct}

An example of full-band maps for ALMA Band~3 is put into context with co-aligned images from SDO in Fig.~\ref{fig:almasdo}. 
The statistics for the brightness temperature values for the whole time series are provided in Table~\ref{tab:almatb}.
The observed Quiet Sun region contains a few magnetic network elements that are mostly located in the left and top of the FOV. 
The network elements appear brighter and thus hotter than their surrounding in the ALMA map (see also the HMI magnetogram in Fig.~\ref{fig:almasdo}c). 
A corresponding network mask (which excludes the outermost part of the FOV) is marked in Fig.~\ref{fig:almahist}a.  
Most of the remaining FOV is characterised by a dynamic mesh-like pattern resembling the pattern seen in other chromospheric diagnostics \citep[e.g.][]{2006A&A...459L...9W}. 
The pattern contains dark regions although their temperature differences with respect to the immediate surrounding varies a lot. 
Occasionally, elongated features become discernible temporarily and  remind of parts of fibrils or dark compact arches 
but the visibility of these features varies in time\footnote{The features are better visible in the movies provided as online material.}. 
A comparison of the ALMA maps with the SDO maps suggests that at least some of these features might be connected to extended magnetic loops as most notably seen in the 17.1\,nm map (Fig.~\ref{fig:almasdo}e and Fig.~\ref{fig:almaoverview}a). 
These features might therefore be caused by weak opacity contributions from coronal loops in the line of sight that then result in weak imprints in the ALMA maps. If and how significant this effect is should be investigated in the future. 
The correlation between the ALMA and SDO maps is discussed in Sect.~\ref{sec:almasdocorr}. 
The small region marked with a red circle in Fig.~\ref{fig:almasdo} is discussed in detail in Sect.~\ref{sec:almaloops}.

\begin{figure*}[tp!]
\sidecaption
\includegraphics[width=12cm]{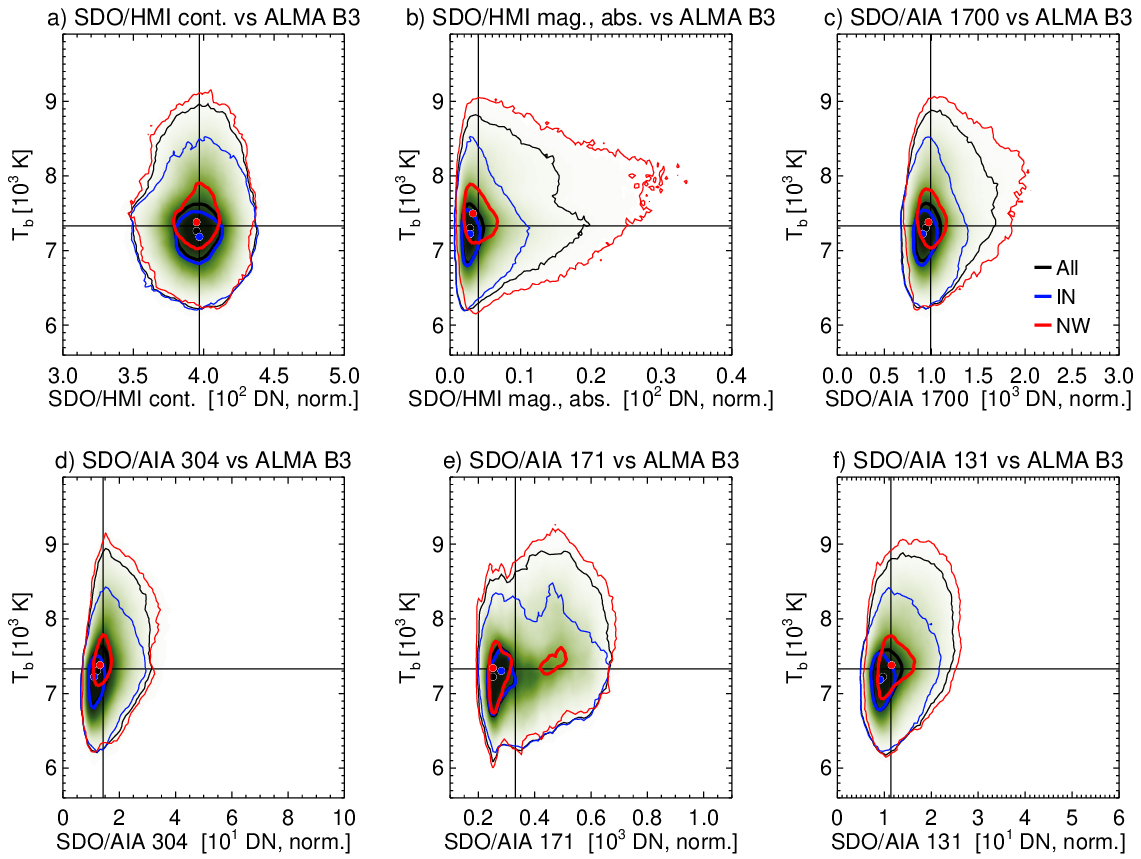}
\caption{Correlation of ALMA Band~3 (full-band) with selected SDO channels: 
\textbf{a)}~HMI continuum,   
\textbf{b)}~HMI magnetogram,   
\textbf{c)}~AIA\,1700,    
\textbf{d)}~AIA\,304,   
\textbf{e)}~AIA\,171,  and
\textbf{f)}~AIA\,131. 
All pixels in the inner region of the FOV (excluding the region with the compact loops) for the whole observing period are considered. 
All SDO maps are convolved with the representative ALMA Band~3 beam. 
The two-dimensional histograms in each panel are presented with green shades for all pixels and also with contour levels for  all pixels (black), internetwork (blue) and network (red).  
Thick contours mark levels of 0.5 and thin contours level of 0.01 with respect to the maximum histogram value. 
The coloured dots show the corresponding maxima. 
The straight black lines represent the median value for ALMA~Band~3 (all pixels) and for the selected SDO channel.
Please note that the SDO data  used here are integration-time-corrected level~1 data numbers (DN) on a linear scale.}
\label{fig:sdocorr}
\end{figure*}

\subsection{Brightness temperature distribution}
\label{sec:tbdistribut}

\begin{table}[b]
    \caption{Observed brightness temperatures in ALMA Band~3. Only the inner region of the FOV within a radius of 22\,\arcsec{} is considered. Separate values are given for all considered pixels and for the subsets marked as network (NW) and internetwork (IN). Compare Fig.~\ref{fig:almahist}a for the pixel map.  }
    \label{tab:almatb}
    \centering
    \begin{tabular}{l|r|r|r}
\hline
&&&\\[-2mm]
Quantity & All &NW&IN\\
&&&\\[-2mm]
\hline
&&&\\[-2mm]
avg.            &   7400 & 7588 & 7228\\
median          &   7363 & 7533 & 7223\\
RMS             &    453 &  478 &  342\\
min.            &   5814 & 5828 & 5814\\
max.            &   9690 & 9690 & 9678\\
1st perc.       &   6515 & 6565 & 6487\\
99th perc.      &   8690 & 8814 & 8195\\
histogram, max. &   7325 & 7415 & 7254\\
histogram, FWHM &    752 &  782 &  702\\              
\hline
\end{tabular}
\vspace*{2mm}
\end{table} 

Taking into account the whole observation sequence with all 1200~time steps, results in brightness temperatures ranging from  $\sim$4440\,K to  $\sim$10700\,K for all pixels and $\sim$5810\,K to  $\sim$9690\,K in the inner region \mbox{($r \leq 22$\,\arcsec)}.  
The corresponding average and standard deviation is $(7500 \pm 514)$\,K for the whole FOV and $(7400 \pm 453)$\,K for the inner region. 
The brightness temperature distribution for the whole observing period for the inner region has a maximum at 7325\,K (see Fig.~\ref{fig:almahist}b). 
In addition, distributions are shown separately for internetwork and network pixels in Fig.~\ref{fig:almahist}b, respectively (see Fig.~\ref{fig:almahist}a for the pixel mask). 
The distribution for network pixels has a peak at a higher temperature compared to the internetwork distribution and deviates clearly from a Gaussian distribution, exhibiting a stretched tail at higher temperatures.  
The values for the maxima and widths of the distributions are provided in Table~\ref{tab:almatb}. 
The difference of the distribution maxima for the network pixels and the internetwork pixels is 160\,K, whereas the average temperature differs by 360\,K. 
The distributions for the network is about 11\,\% broader than for the internetwork except with a FWHM values of 782\,K and 702\,K, respectively. These results are compared to other ALMA observations in Sect.~\ref{sec:disctemp}.
%

\subsection{Temporal variation}
\label{sec:tbtimevar}

One position in the internetwork and one in the network are selected and marked in Fig.~\ref{fig:almahist}a and labelled A and B, respectively. 
The temporal evolution of the brightness temperature at these locations is shown in Fig.~\ref{fig:almaprof}a-b.
There seems to be an oscillation with a period on the order of 3\,min as it is expected for chromospheric internetwork regions although such variations are more pronounced for other locations in the internetwork.
A more detailed study of the oscillatory behaviour will be published in a forthcoming paper \citep{jafarzadeh_waves_inprep}.
The network position does not show an equally clear oscillation but variations on different time scales.  
Brightness temperature profiles along the x-axis for the two selected position are shown in Fig.~\ref{fig:almaprof}c and d, respectively. 
The major and minor axes (FWHM ) of the synthetic beams  (see the top of the panels) limit the smallest scales over which variations can be recovered.

\begin{figure*}[tp!]
\includegraphics[width=\textwidth]{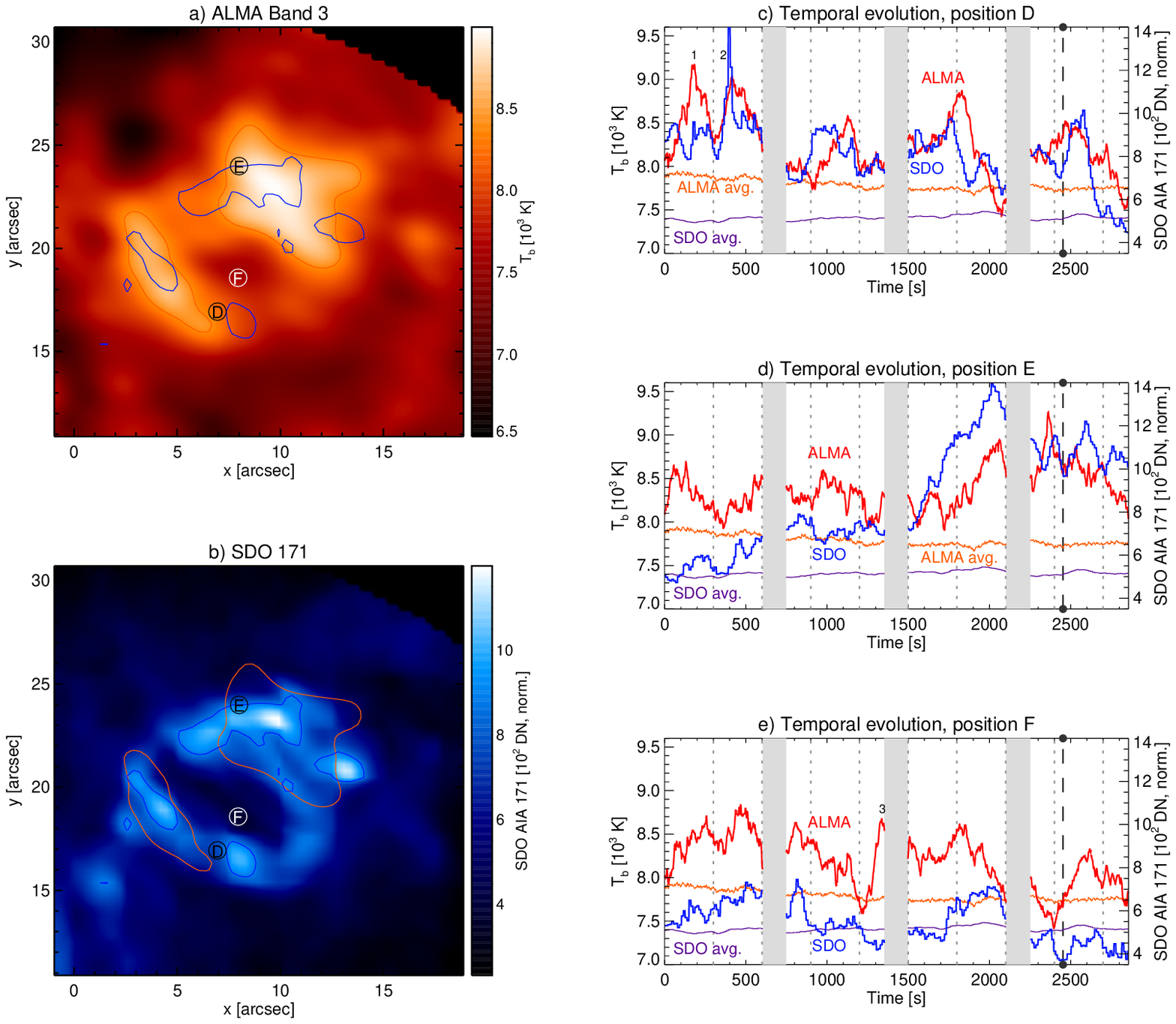}
\caption{Close-up of a group of compact loops (see red circle Fig.~\ref{fig:almasdo}). 
\textbf{a)}~ALMA Band~3 brightness temperature map with three selected positions (D, E, F). The read contours mark a brightness temperature of 8400\,K, whereas the blue contour encloses SDO\,AIA171 values of 880\,DN. 
\textbf{b)}~Corresponding SDO\,AIA171 map with the same contours as in panel~a and   
\textbf{c-e)}~Time evolution for the Band~3 temperatures (thick red lines) and the the SDO~171 values (thick blue lines) for the three selected positions (D: black, E: red, F: blue). 
Three temperature rise events are marked with small numbers (1-3). 
For comparison, the average over all pixels contained in the close-up region are shown as orange lines for ALMA
and purple lines for SDO~171.  
The calibration breaks with no science data are marked as grey shaded areas.
Please note that the SDO data  used here are integration-time-corrected level~1 data numbers (DN) on a linear scale.}
\label{fig:loops}
\end{figure*}

\subsection{Correlation of ALMA and SDO maps}
\label{sec:almasdocorr}
 
A comparison of the ALMA  maps with the SDO maps reveals that the extended magnetic loops as most notably seen in the 17.1\,nm map (Fig.~\ref{fig:almasdo}e) also leave weak imprints in the ALMA maps.  
In order to quantify such similarities, the cross-correlations of the ALMA maps with the corresponding  SDO maps are calculated for the considered SDO channels.
For each time step, the SDO maps are first convolved with the representative ALMA Band~3 beam before calculating the cross-correlation for the whole inner FOV region ($r < 22\arcsec$) but excluding the circular region with the compact loops (red circles in Fig.~\ref{fig:almasdo}). 
The resulting time-averaged values $\langle\mathcal{C}\rangle_t$ are highest for  ALMA~Band~3 -- SDO/AIA~30.4\,nm and  ALMA~Band~3 -- SDO/AIA~13.1\,nm, both  reaching a moderate correlation of \mbox{$\langle\mathcal{C}\rangle_t=0.38$},  
followed by SDO/AIA~17.1\,nm with $\langle\mathcal{C}\rangle_t = 0.33$.
The cross-correlation values for only network pixels in the inner region are $\langle\mathcal{C}\rangle_t = 0.35$, 0.34, and 0.33 for SDO~AIA~17.1\,nm, SDO/AIA~30.4\,nm, and SDO/AIA~13.1\,nm, respectively. 
In general, the correlation is much weaker for internetwork pixels with values staying below 0.28 (SDO/AIA~30.4\,nm) and 0.19 (SDO/AIA~13.1\,nm). 
The cross-correlation with selected SDO channels is visualized in Fig.~\ref{fig:sdocorr} for all pixels and also for network and internetwork pixels separately. 
The plots for SDO/AIA~17.1\,nm and SDO/AIA~13.1\,nm (panels e-f) reveal the tendency of increasing brightness temperature with increasing SDO count value, implying that statistically a higher value in these channels is connected to a higher brightness temperature along the same line of sight.  
Please refer to Sect.~\ref{sec:disc_height} for a discussion of potential implications for the formation height ranges of  ALMA 
Band~3.

\subsection{Compact loops}
\label{sec:almaloops}
 
In the top right of the interferometric FOV, a group of short magnetic loops is visible in most SDO channels (see encircled region in Fig.~\ref{fig:almasdo} and   Fig.~\ref{fig:loops}b for a close-up).
The loops connect patches of opposite polarity as visible in the HMI magnetogram. 
Given the appearance in the SDO maps, we suggest that the features in the ALMA maps  (see Fig.~\ref{fig:loops}a) are unresolved loop strands. 
The projected lengths of the loops  are on the order of 10\,\arcsec, which agrees with the distance between the magnetic foot points seen in the HMI map (Fig.~\ref{fig:almasdo}b). 
The ALMA map shows higher brightness temperatures at roughly the same location as the hotter SDO channels (panels~d-f), implying that ALMA maps the hot loop tops at brightness temperatures between 8500\,K and a maximum of 9650\,K. 
%
Several elongated features with enhanced temperature are discernible. Their widths are with 2-3\,\arcsec{} close to the resolution limit, whereas the distance of 4-6\,\arcsec{} between the elongated features is resolved. 
Between these features, the brightness temperature can be as low as 7500\,K and thus close to the average value for the whole FOV. 
The brightness temperature at the loop tops varies strongly in time between around 8000\,K and peak values of close to or in excess of 9000\,K. 
The temperature difference between the loop tops and the surrounding is thus often on the order of 1000\,K or more although it varies, resulting in varying contrast of the loops. 
The brightness temperature variations are compared to the corresponding variations in SDO~AIA171 for three selected positions (D, E, F) in \mbox{Fig.~\ref{fig:loops}c-e.}
At the beginning of the observing period, some loop tops exhibit several consecutive peaks with about 4-5\,min in-between, which may imply  oscillatory behaviour. 
Positions D and E show are located on the loops whereas position~F marks a cooler region in-between the loops for the shown time step in Fig.~\ref{fig:loops}a.
It is quite clear that loops are not properly resolved and also sway in time, thus affecting the signal at a given fixed spatial position. 
The loop position~D shows a strong temperature rise from 8000\,K at $t = 20$\,s to 9200\,K at $t = 170$\,s, i.e. 1200\,K over 150\,s with a corresponding rate of $\sim$8\,K\,s$^{-1}$, here referred to as event~1. 
This event is followed by another temperature rise (event~2) with a rate of $\sim$6\,K\,s$^{-1}$.  
For event~2, the SDO\,AIA171 signal also steeply increases at the same time
whereas there is only a moderate increase for   event~1. 
The changes in SDO~AIA171 is not always tightly coupled to the changes in ALMA brightness temperature as is most obvious for event~3 at position~F. 
For that event, a steep temperature rise of 1100\,K over 120\,s is observed (rate: $\sim$9\,K\,s$^{-1}$) whereas the SDO~AIA171  signal slowly decreases.

Analysis of the SDO data for the interferometric FOV and the surroundings (see Fig.~\ref{fig:almaoverview}) before and after the ALMA observation, suggests that the compact loop system is the result of a flux emergence event. 
The first loop top appears in HMI magnetograms around UT9:11, i.e. about five hours prior to the ALMA observation. 
Subsequently, two footpoints with opposite magnetic polarity move away from each other and reach their final separation within one hour. In that process, further footpoints emerge next to the initial ones and finally form the group of compact loops. The loops become visible in AIA\,171  maps during that emergence phase. 
After the ALMA observation, the two polarities move towards each other until they mix around UT17:40, followed by the disintegration of the loops. 
The AIA\,304 and AIA\,171 data show that the group finally vanishes from about 20UT.

Such emerging magnetic loops are expected to be optically thin at millimeter wavelengths and may reveal the atmosphere underneath, which could provide the thermal properties of the inner part of emerged regions (Nóbrega-Siverio, priv.comm.). Quite opposite to this expectation, the ALMA observation presented here clearly feature bright magnetic loops, suggesting that they are  optically thick and thus block the view at the possibly existing cool plasma below. 
A possible explanation is that the observed loop tops stay at rather low altitude, at least during the 45\,min covered by ALMA. 
The extended coronal loops that traverse the ALMA FOV (see Fig.~\ref{fig:almasdo}) seem to be located higher in the atmosphere and might prevent the compact loops from rising higher. 
The consequence would be that the latter remain in the chromosphere with 
loops containing plasma with higher density and correspondingly larger opacity.

\section{Discussion} 
\label{sec:disc}
\subsection{Brightness temperature distribution}
\label{sec:disctemp}

The average brightness temperatures in the ALMA Band~3 full-band maps discussed here are on the order of 7400\,K for all pixels in the inner parts of the FOV and on the order of $\sim7590$\,K and $\sim7230$\,K when separating network and internetwork pixels (see Table~\ref{tab:almatb}). 
Accordingly, the difference between the average full-band network and internetwork brightness temperatures is $\sim$360\,K.
As expected, these values are close to the reference value of 7300\,K suggested by \citep{2017SoPh..292...88W} because the absolute brightness temperature scale was corrected accordingly. 
It should be noted, however, that the applied correction was a minor one 
(see Sect.~\ref{sec:observbeam}). 
In the following, we compare the brightness temperature distributions of  the data presented here to the ALMA Band~3 observations from Cycle~4 by  \citet{2019ApJ...877L..26L} and  \citet{2018A&A...619L...6N}.

\citet{2019ApJ...877L..26L} analyse data obtained on April 27, 2017, for a Quiet Sun region at 200\,\arcsec{} distance from solar disk-centre that contains magnetic network and internetwork patches. 
They state a width  of 1.6\,\arcsec\, for both axes of their synthesized beam, which is slightly smaller than the representative beam used for the data presented here.   
The brightness temperatures range from 5630 K to 9140\,K in their time-averaged map and from 4370\,K to 11170\,K in the corresponding time sequence at 2\,s cadence. 
The lowest temperatures are found in a 20\,\arcsec{} wide region, which is significantly cooler than the surrounding atmosphere but which is not visible at other wavelengths as observed with SDO. 
From their Fig.~2, we determine the peaks and FWHM values of the brightness temperature distributions for the selected network and internetwork patches. 
The network patch has a maximum at  7340\,K and a FWHM  of 1240\,K, whereas the internetwork regions have maxima at 7200\,K and 7090\,K and FWHM s of 530\,K and 430\,K and corresponding standard deviations of 225\,K and 183\,K, respectively.
%
In contrast, their cool region has a maximum at 6330\,K and a FWHM  of 1470\,K. 
%
%
The distribution peak temperatures for the internetwork regions are only slightly lower than found in this study  (see Table~\ref{tab:almatb} and Sect.~\ref{sec:tbdistribut}) but it might be argued that they nonetheless agree within the expected uncertainties of possibly a few 100\,K. 
The FWHM of the internetwork temperature distribution found by  \citet{2019ApJ...877L..26L} is significantly smaller than for the data set analysed here (702\,K, see Table~\ref{tab:almatb}). 
Their distribution for network pixels, on the other hand, has a maximum at 75\,K lower than the value found here see Table~\ref{tab:almatb}) but it still agrees within the error limits. 
The corresponding FWHM, however, is much larger than found in this study. 
As we will demonstrate in Sect.~\ref{sec:angres}, the FWHM of the brightness temperature distribution depends on the effective angular resolution of the observation and thus on a number of factors ranging from seeing conditions to details of the image reconstruction procedure. 

\citet{2018A&A...619L...6N} observed the Sun with ALMA in Band~3 on  March 16, 2017 for several positions from the limb to disk-center. 
For the latter, they found an average brightness temperatures of 7530\,K for network pixels, 6940\,K for internetwork (cell) pixels, and 7220\,K as average over the FOV. 
We find that the average brightness temperatures for network pixels agree quite well with values for the observations presented here whereas the value for internetwork pixels  found by \citeauthor{2018A&A...619L...6N} is almost 300\,K lower than the value found here.  
Accordingly, they state an average difference between network and internetwork of 590\,K, whereas it is only 360\,K in the data presented here (see Table~\ref{tab:almatb}). 
Furthermore,  \citet{2018A&A...619L...6N} determined the standard deviation over the FOV as 390\,K as compared to 
$\sim 450$\,K for all pixels in the inner region of the data presented here. 
The values found by \citeauthor{2018A&A...619L...6N} is similar to the value found here for internetwork pixels ($\sim$340-400\,K) and lower than the corresponding network value ($\sim$500\,K). 
\citeauthor{2018A&A...619L...6N} state that the synthetic beam of their disk-center observation has a major axis of 8.1\,\arcsec{} and a minor axis of 2.3\,\arcsec, which is significantly larger than the synthetic beams for the data presented here. 
Please note that \citeauthor{2018A&A...619L...6N} achieved smaller beams for earlier observations closer to the solar limb. 
The differences in brightness temperatures between those found by \citeauthor{2018A&A...619L...6N} at solar disk-centre and those reported here might therefore be partially due to the differences in angular resolution in addition to differences arising from the applied post-processing method. 
We also note that the data run used by \citeauthor{2018A&A...619L...6N} was obtained under worse seeing conditions with a higher amount of precipitable water vapour (PWV) in Earth's atmosphere. 

For comparison, we considered the  BIMA observation by \citet{2006A&A...456..697W} at 85\,GHz with a beam (and thus an angular resolution) of 10\,\arcsec. 
\citeauthor{2006A&A...456..697W} find rms variations of $\sim$120\,K for both network and interwork locations. This value is about a factor 3-4 less than for the ALMA results discussed above.

\begin{figure}[tp!]
\includegraphics[width=\columnwidth]{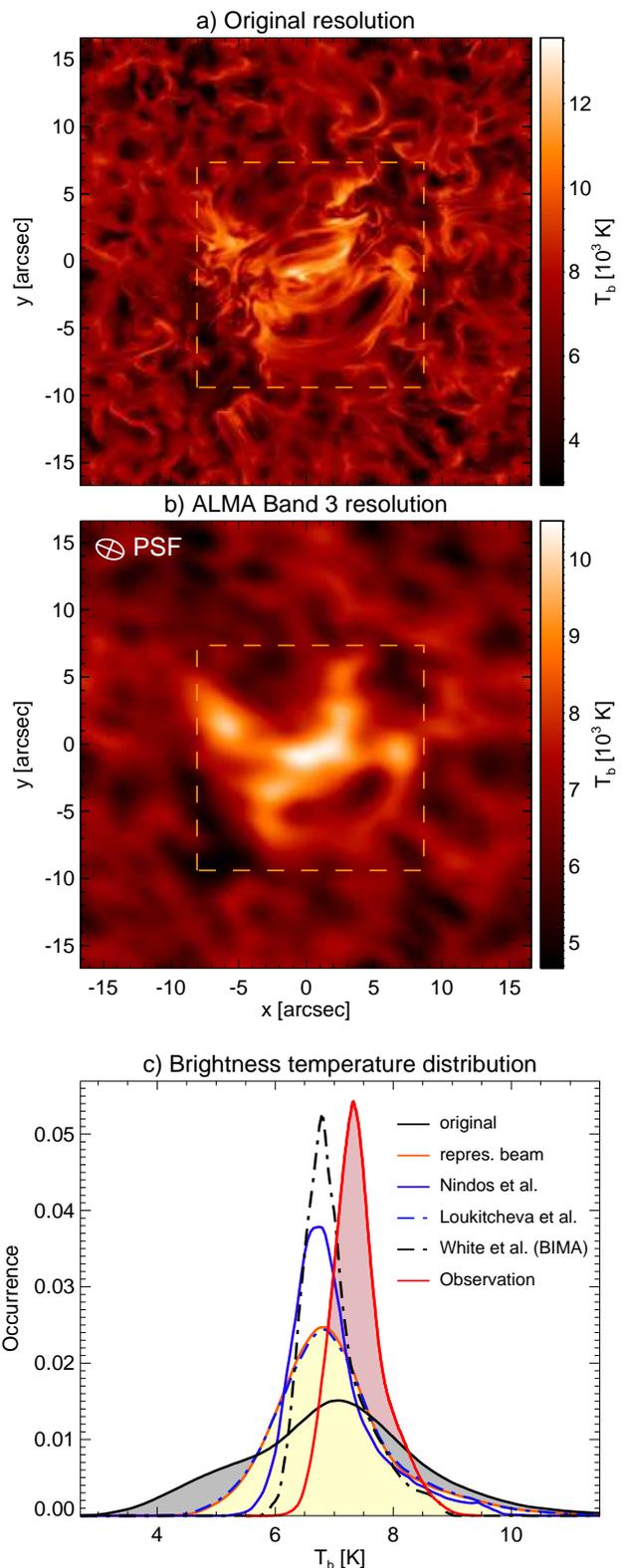}
\vspace*{-9mm}
\caption{Selected synthetic brightness temperature map (Bifrost/ART)  for the average Band~3 frequency (100\,GHz). The enhanced network region is marked with a yellow dashed rectangle. 
\textbf{a)}~The original mm map averaged over the frequencies covered by the solar ALMA observations in Band~3. 
\textbf{b)}~The mm map after applying the PSF (see upper left corner) corresponding to the representative synthetic beam for the observation presented here. 
\textbf{c)}~Brightness temperature distributions for the original map (black solid line, grey area) and after applying different PSFs: representative beam as in panel~b (orange solid, yellow area), \citet{2018A&A...619L...6N} (blue solid), \citet{2019ApJ...877L..26L} (blue dot-dashed), and \citet{2006A&A...456..697W} (BIMA, black dot-dashed).
The distribution for observed Band~3 temperatures is plotted as red line and red shaded area (all pixels, see Fig.~\ref{fig:almahist}b).}
\label{fig:simres}
\end{figure}
\begin{table}[b]
    \caption{Simulated brightness temperatures in ALMA Band~3 after applying different synthetic beams in comparison to the ALMA observations. The average and standard deviation (rms) are given for all pixels and separately for network (NW) and internetwork (IN) pixels. The major and minor axes (and if applicable the angle) of the applied beam are stated for each case.} 
    \label{tab:simtb}
    \centering
    \begin{tabular}{l|r|r|r|c|c|c}
\hline
&&&&\multicolumn{3}{c}{Beam}\\
Quantity & All &NW&IN&maj&min&ang\\
\hline
\multicolumn{7}{c}{}\\[-3mm]
\multicolumn{7}{c}{Original model}\\[1mm]
\hline
$T_\mathrm{b}$, avg. [K]&   7015    &   7977    &   6688  &--&--&--\\
$T_\mathrm{b}$, rms  [K]&   1549    &   1794    &   1304  &&&\\
\hline
\multicolumn{7}{c}{}\\[-3mm]
\multicolumn{7}{c}{Convolved with representative ALMA beam, this work.}\\[1mm]
\hline
$T_\mathrm{b}$, avg. [K]&   7015    &   7969 &  6691 &2.10\arcsec  &1.37\arcsec& 68.0$^\circ$\\
$T_\mathrm{b}$, rms  [K]&   1033    &   1254 &   693 &&&\\
\hline
\multicolumn{7}{c}{}\\[-3mm]
\multicolumn{7}{c}{Convolved with ALMA beam by Loukitcheva et al. (2019).}\\[1mm] 
\hline
$T_\mathrm{b}$, avg. [K]&   7015    &   7971 &  6690 &1.6\arcsec&1.6\arcsec&--\\
$T_\mathrm{b}$, rms  [K]&   1056    &   1279 &   722 &&&\\
\hline
\multicolumn{7}{c}{}\\[-3mm]
\multicolumn{7}{c}{Convolved with ALMA beam by Nindos et al. (2018)}\\[1mm]
\hline
$T_\mathrm{b}$, avg. [K]&   7015    &   7865 &  6726 &8.1\arcsec&2.3\arcsec&-48$^\circ$\\
$T_\mathrm{b}$, rms  [K]&   744    &   802 &   431 &&&\\
\hline
\multicolumn{7}{c}{}\\[-3mm]
\multicolumn{7}{c}{Convolved with BIMA beam by White et al. (2006).}\\[1mm]
\hline
$T_\mathrm{b}$, avg. [K]&   7016    &   7681 &  6790 &10.0\arcsec&10.0\arcsec&--\\
$T_\mathrm{b}$, rms  [K]&   536    &   495 &   317 &&&\\
\hline
\multicolumn{7}{c}{}\\[-3mm]
\multicolumn{7}{c}{ALMA observations, this work. }\\[1mm]
\hline
$T_\mathrm{b}$, avg. [K]   &   7400 & 7588 & 7228&2.10\arcsec  &1.37\arcsec& 68.0$^\circ$\\
$T_\mathrm{b}$, rms  [K] &    453 &  478 &  342&&&\\
\hline
\end{tabular}
\end{table}

\subsection{Dependence on angular resolution.} 
\label{sec:angres}

As already demonstrated by \citet{2007A&A...471..977W}, not resolving 
small-scale chromospheric features due to limited angular resolution results in a reduction of the corresponding standard deviation in the obtained brightness temperature maps. 
The better the angular resolution, the higher the standard deviation in the observations. 
In the following, we test the influence of reduced angular resolution on the resulting brightness temperature distribution by convolving  synthetic brightness temperature maps with different synthetic beams.
Brightness temperature maps for ALMA Band~3 frequencies are calculated with the Advanced Radiative Transfer (ART)  code (de la Cruz Rodriguez et al., in prep.) for a time series of snapshots from a  3D radiation magnetohydrodynamic simulation with Bifrost \citep{2016A&A...585A...4C,2011A&A...531A.154G}. 
The series used here has a duration of 20\,min and a cadence of 1\,s and features an enhanced network region in the middle with surrounding Quiet Sun. 
For each time step, the maps for the different frequencies are averaged, resulting in band-average maps.  
See Fig.~\ref{fig:simres}a for an example for a selected time step.  
Applying the representative beam for the data presented here (see Sect.~\ref{sec:observbeam}), produces a 
brightness temperature map at an angular resolution equivalent to the analysed ALMA observations (see Fig.~\ref{fig:simres}b).
This procedure is repeated for all maps in the time series and also for the ALMA beams used by \citet{2018A&A...619L...6N} and  \citet{2019ApJ...877L..26L}, and the BIMA beam by \citet{2006A&A...456..697W}. 
The resulting brightness temperature distributions for the original maps and the degraded maps for all four beams are compared to the observational results in Fig.~\ref{fig:simres}c. 
All time steps are taken into account. 
For the elliptic beams such as in the observations presented here and for \citeauthor{2018A&A...619L...6N},  additional degraded maps are calculated with the beam rotated by $90\,\deg$. 
This extra step reduces possible artificial effects due to the coincidental alignment of elongated features in the original map with a beam axis. 
The resulting averages and standard deviations of the brightness temperature maps are summarised in Table~\ref{tab:simtb}.

The original maps have an average of 7015\,K and a standard deviation of 1549\,K. The network pixels in the middle of the map (see dashed rectangle in Fig.~\ref{fig:simres}a-b) have an almost 1000\,K higher average and a larger standard deviation whereas both are reduced for the internetwork pixels (outer region of the map in  Fig.~\ref{fig:simres}a-b).) 
Reducing the angular resolution by convolution with a synthetic beam (i.e. a PSF) does not affect the brightness temperature average but results in a narrower distribution (see Fig.~\ref{fig:simres}c) and a correspondingly reduced standard deviation (Table~\ref{tab:simtb}).
Using the representative beam from the ALMA observations presented here results in a standard deviation of 1033\,K, which is very similar to the results obtained with the symmetric 1.6\arcsec{} wide beam reported by \citet{2019ApJ...877L..26L}. 
The larger and more elliptical beam by \citet{2018A&A...619L...6N} results in even lower standard deviation of 744\,K. 
For comparison, we also apply the 10\arcsec{} BIMA beam by \citet{2006A&A...456..697W}, which returns a standard deviation of 536\,K for the whole map and 317\,K for internetwork pixels although a substantial mixing of network and internetwork within the beam is expected.

The average brightness temperatures for the whole maps (``all'' in Table~\ref{tab:simtb}) are only 300\,K lower than those derived from the observations presented in this work. The simulated standard deviation, however, is roughly a factor two higher than the corresponding observational value. It is important to note that the original simulated maps represent the best possible maps that can be obtained with a given beam, whereas additional factors can lead to a further reduction of the standard deviation in the observed maps. First of all, interferometric snapshot observations with a finite number of antennas can by nature never provide a truly complete coverage of the spatial Fourier space and resulting degradation must be expected. Furthermore, seeing conditions, noise contributions and technical details of the imaging process itself are possible causes for further reduction. On the other hand, these first results are already very promising.

We conclude that our results agree with the study by \cite{2019ApJ...877L..26L} at least on a qualitative level and also in some aspects with \citet{2018A&A...619L...6N} but more systematic statistical comparisons should be attempted in the future.
There are many factors that influence the brightness temperature distribution ranging from the properties of the observed target regions and accuracy of the applied network mask to different seeing condition and details of the imaging procedure. 
The small size of the FOV and thus the peculiarities of the observed regions will produce variations in the  statistical properties derived from different observations. Such results should be compared to a corresponding analysis of  mosaicking data that cover larger FOVs \citep[see, e.g.,][]{2017ApJ...845L..19B,2019A&A...622A.150J}.
Furthermore, the test for different angular resolutions implies that more extended array configurations of ALMA, which might be offered in future observing cycles, are likely to lead to higher rms variations and thus more contrast in the reconstructed images.


\subsection{Formation height} 
\label{sec:disc_height}
\citet[][see also references therein]{2017SoPh..292...88W} point out that contributions from the corona to brightness temperatures measured with ALMA should be expected and that the contributions could amount to a few 100\,K in Band 3 from the densest parts of the corona. 
As mentioned in Sect.~\ref{sec:observstruct} and quantified in terms of cross-correlations in Sect.~\ref{sec:almasdocorr}, coronal loops that extend across the ALMA field of view and are clearly visible in coronal SDO channels can leave very weak imprints in some ALMA Band 3 maps but are best seen in movies. 
Internetwork and network regions are clearly seen in the ALMA maps presented here but appear to be more horizontally expanded than SDO/AIA~170\,nm maps, which, together with the rather weak to moderate cross-correlation between this SDO channel and ALMA Band~3, may imply that Band~3 is formed above the layer from where the continuum radiation at 170\,nm emerges. 
At the same time, one should be cautious with concluding on the formation height range based on these arguments, especially regarding the cross-correlations, even with these findings supporting the claim by \citet{2017SoPh..292...88W}. 
Rather, it is essential to study  the mapped height ranges and contribution functions along the line of sight in ALMA data in detail. 
The scientific potential of the measured brightness temperatures can only be truly unfolded once the temperatures can be assigned to precise height ranges and thus being translated into measurements of the chromospheric temperature stratification.

\section{Conclusion and Outlook} 
\label{sec:conc}

Although the solar observing mode of ALMA is still in its early development phase, the ALMA Band~3 data presented here and in previous publications already demonstrate ALMA's potential for scientific studies of the solar chromosphere. 
The spatial resolution currently achieved in Band~3 certainly limits the study of the chromospheric small-scale structure and dynamics but, at the same time, and this cannot be emphasized enough, it is an enormous leap forward for the observation of the Sun at millimeter wavelengths. 

With this tool at hand, the brightness temperature distribution for a Quiet Sun region at disk-centre is quantitatively analysed, also separated in network and internetwork patches, and can thus serve as important test for numerical simulations of the solar atmosphere.

While many aspects such as the exact formation height ranges, possible weak coronal contributions, and details of the imaging procedure need to be investigated in more detail, the presented data already allows for a large range of scientific studies. 
For instance, we are able to measure the brightness temperatures in a group of compact loops as function of time.

\section*{Acknowledgments}
This work is supported by the SolarALMA project, which has received funding from the European Research Council (ERC) under the European Union’s Horizon 2020 research and innovation programme (grant agreement No. 682462), and by the Research Council of Norway through its Centres of Excellence scheme, project number 262622.
JdlCR is supported by grants from the Swedish Research Council (2015-03994), the Swedish National Space Board (128/15) and the Swedish Civil Contingencies Agency (MSB). This project has received funding from the European Research Council (ERC) under the European Union's Horizon 2020 research and innovation programme (SUNMAG, grant agreement 759548). The Institute for Solar Physics is supported by a grant for research infrastructures of national importance from the Swedish Research Council (registration number 2017-00625).
This paper makes use of the following ALMA data: ADS/JAO.ALMA\#2016.1.00423.S. ALMA is a partnership of ESO (representing its member states), NSF (USA) and NINS (Japan), together with NRC(Canada), MOST and ASIAA (Taiwan), and KASI (Republic of Korea), in co-operation with the Republic of Chile. The Joint ALMA Observatory is operated by ESO, AUI/NRAO and NAOJ. We are grateful to the many colleagues who contributed to developing the solar observing modes for ALMA and for support from the ALMA Regional Centres. 
We acknowledge support from the Nordic ARC node based at the Onsala Space Observatory Swedish national infrastructure, funded through Swedish Research Council grant No 2017 – 00648, and collaboration with the Solar Simulations for the Atacama Large Millimeter Observatory Network (SSALMON, http://www.ssalmon.uio.no).
The ISSI international team 387 ``A New View of the Solar-stellar Connection with ALMA''  was funded by the International Space Science Institute (ISSI, Bern, Switzerland). 
We thank D.~E.~N\'obrega Siverio for helpful comments regarding the observed compact loops.


\bibliographystyle{aa}
\bibliography{swmain.bib}
\appendix
\section{Introduction to interferometric observations} 
\label{sec:appendix}
\subsection{Interferometric image reconstruction} 
\label{sec:app_intimg}

Reconstructing solar images from interferometric observations is by its very nature a challenging task that is limited by the achieved sampling of the spatial Fourier domain. 
Each measurement made by a pair of antennas (a ``baseline'') with a particular spacing (measured in wavelengths) and orientation corresponds to a single Fourier component\footnote{The spatial Fourier space is also referred to as `` uv space''. A component in the uv space is determined by the separation of the two involved antennas (i.e. the baseline length), the observing frequency, and the angles under which the source is observed on the sky.}
of the source brightness distribution. 
The more Fourier components measured, the more reliable the original source image can be reconstructed from the available information.  
For sources that do not change significantly during an observation, the Earth's rotation can be utilized to greatly increase the number of Fourier components measured owing to the fact that the orientation of the antenna distribution relative to the source changes in time. This is referred to as "Earth rotation aperture synthesis".
While this technique can be used to better sample the Fourier domain, it is not applicable in the case of variable sources like the Sun. 
It could be argued that using measurements over short time periods is a justifiable compromise between quality of the reconstructed images and time resolution. 
The common procedure for the studies published so far \citep[see, e.g.,][]{2018A&A...619L...6N,2019ApJ...877L..26L} is therefore to reconstruct single images from all available frequency channels from all sub-bands over extended time windows of up to an entire scan ($\sim 10$ min) or even the full duration of the observations ($\sim 1$ hr). 
We refer to maps that are constructed from all frequency channels as ``full-band'' maps, regardless the time window used, whereas the term ``snapshot'' images is used for maps that are made every integration time. The latter would result in time series of maps with 2\,s cadence for Cycle~4 data.

As the Fourier component sampled by a certain baseline also depends on the observing frequency, the sampling in the Fourier domain can be improved by using larger frequency ranges for the imaging process, combining spectral channels and or even all sub-bands.  
This technique, referred to as "frequency synthesis", essentially averages out information in the spectral domain in favour of better image fidelity. Obviously, care must be taken in evaluating the trade-offs between improved sampling - and the possibility of improved imaging fidelity - and the loss of information caused by averaging over temporal or spectral variations. 
Depending on the intended scientific aim, this might not be desirable and image reconstruction resulting in slightly less reliable brightness temperatures might be acceptable in return for access to the spectral domain. 
The reliability of brightness temperature differences after splitting into sub-bands is under investigation and will be discussed in greater detail in future publications.
For this first paper, we produce maps using all four sub-bands (full-band maps) but we do not average the data in time.

\subsection{Interferometric field-of-view} 
\label{sec:app_fov}

The FOV of the interferometric observations is set by the primary beam, which is due to the aperture of a single antenna. The effect of the primary beam response is to multiply the field of view by an approximately Gaussian function, referred to as the primary beam taper. 
The size of the Gaussian primary beam is typically specified in terms of its  full-width-at-half-maximum (FWHM), which depends on the observed wavelength \citep[or frequency, see, e.g.,][]{2016SSRv..200....1W}.

\subsection{Synthesized beam} 
\label{sec:app_beam}

\begin{figure}[tp!]
\begin{center}
\includegraphics[width=\columnwidth]{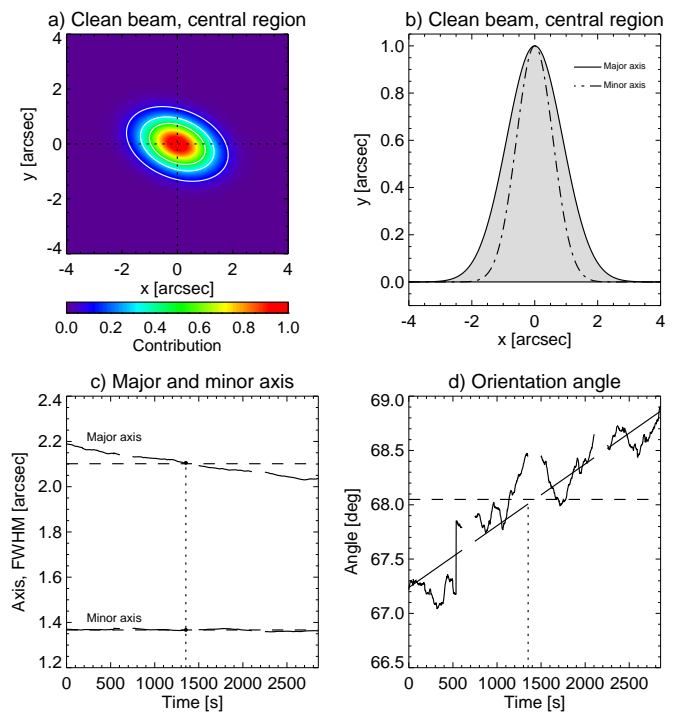}
\end{center}
\caption{The synthesized beam (here also referred to as point spread function, PSF) representative for the Band~3 solar observations. The beam corresponds to the central frequency of 100\,GHz. 
\textbf{a)}~Coloured contour plot for the central region around the main lobe, \textbf{b)}~profiles along the major and minor axes, \textbf{c)}~variation with time for the major and minor axes, and \textbf{d)}~for the orientation angle. 
The dashed horizontal lines in the lower panels mark the time-averages of the beam parameters, while the dotted vertical lines mark the time of the  actual full-band beam that is closest in size to the representative beam as derived from the time-average of the major and minor axes.}\label{fig:almapsf}
\end{figure}

The synthesized (interferometric) beam, which corresponds to the point spread function (PSF) of the interferometric array, is calculated during the image reconstruction process and provided as output. 
The PSF depends on the angles under which the target appears on the sky and thus changes with time when the target is tracked during an observation. The ALMA PSF is characterized by a central lobe normalized to unit amplitude and sidelobes due to incomplete sampling in the Fourier domain. The central lobe is approximately an elliptical Gaussian, whose dimensions are determined by the maximum distances between antennas, thus determining the angular resolution of the observation. Image deconvolution involves removal of the PSF sidelobe response from the image leaving, in principle, a map that is the true brightness distribution convolved with the "clean beam" plus noise.  
In general, the (clean) beam is simply the elliptical Gaussian fit to the central lobe of the PSF (Fig.~\ref{fig:almapsf}a).

\subsection{Absolute temperatures based on Total Power maps}
\label{sec:app_tp}
The interferometric observations provide brightness temperature differences relative to the mean brightness but lack an absolute offset corresponding to the measurement at the zero spatial frequency (or zero-spacing frequency) in the Fourier domain. 
That is, the zero component corresponds to a telescope pair with a baseline of zero length. 
Since this is technically impossible, ALMA  overcomes this problem by combining interferometric data with observations with single-dish total power antennas.  
These antennas have a diameter of 12\,m like the other antennas in the ALMA's 12-m array. The angular resolution of the full disk maps produced by TP fast-scanning techniques is therefore identical to the FOV of the 12-m array. 
While details on spatial scales below $\approx 60$\,\arcsec{} remain unresolved in TP maps, the features in the map can clearly be correlated with full-disk maps obtained with the Solar Dynamics Observatory  \citep[SDO;][]{2012SoPh..275....3P} (see Sect.~\ref{sec:coobserv}).
Note that a single measurement with a TP antenna toward the interferometric target provides the zero-spacing component in Fourier space, whereas fast-scanning over the solar disk measures the brightness distribution of the source on angular scales of the primary beam (here $\sim 60$\,\arcsec) and larger (to the angular extent of the Sun itself).

\end{document}